\documentclass[titlepage,12pt]{article}
\usepackage{amssymb,psfig,epsfig}
\newcommand{\Li}[1]{\mbox{Li}_2\left(#1\right)}
\newcommand{\bea}{\begin{eqnarray}}
\newcommand{\eea}{\end{eqnarray}}
\newcommand{\be}{\begin{equation}}
\newcommand{\ee}{\end{equation}}

\newcommand{\nn}{\nonumber}

\newcommand{\msbar}{{\overline{\mbox{MS}}}}
\newcommand{\xb}{\bar x}

\newcommand{\xgc}{x_{\rm cut}}
\renewcommand{\Re}{\mbox{Re}}

\renewcommand{\thefootnote}{\fnsymbol{footnote}}

\makeatletter

\begin{document}

%
%
\begin{titlepage}
\noindent
hep-ph/9904251\hfill PITHA 99/10
\vspace{0.8cm}
\begin{center}
{\bf\LARGE 
Production of top quark pairs in association } \\
{\bf\LARGE
with a hard gluon to order $\alpha_s^2$\\}
\vspace{2cm}
\centerline{Arnd Brandenburg\footnote{Research supported by
BMBF, contract 057AC9EP.}$^,$\footnote{
Address after April, 1, 1999: DESY Theory Group, D-22063 Hamburg.}}
\vspace{1cm}
\centerline{Institut f\"ur Theoretische Physik,
RWTH Aachen, D-52056 Aachen, Germany}
\date{\today}
\vspace{3cm}
{\bf Abstract:}\\[2mm]
\parbox[t]{\textwidth}
{
The production of top quark pairs accompanied by a hard gluon
in $e^+e^-$ annihilation is studied including  
next-to-leading order corrections in the strong coupling. 
At leading order, the fraction $r$ of $t\bar{t}g$ events 
with respect to all $t\bar{t}$ events is computed analytically
as a function of the minimal gluon energy. 
Next-to-leading order results for $r$ are given 
for center-of-mass energies of 0.5 and 1 TeV.
We further calculate the  differential distribution of $r$ 
with respect to several variables, 
including the top quark energy and 
the $t\bar{t}$ invariant mass. 
We then investigate how
our results depend on the choice 
of the renormalization scheme for the top quark mass  
by comparing results expressed in terms of either the pole mass 
or the $\msbar$ mass. Finally we estimate the sensitivity of the
fraction $r$ on the value of the running top quark mass at a scale
of 1 TeV.  
}
\end{center}
\end{titlepage}
\newpage
\setcounter{footnote}{0}
\renewcommand{\thefootnote}{\arabic{footnote}}
\setcounter{page}{1}
%
%

\begin{figure}[ht]
\unitlength1.0cm
\begin{center}
\begin{picture}(8,8)
\put(0,0){\psfig{figure=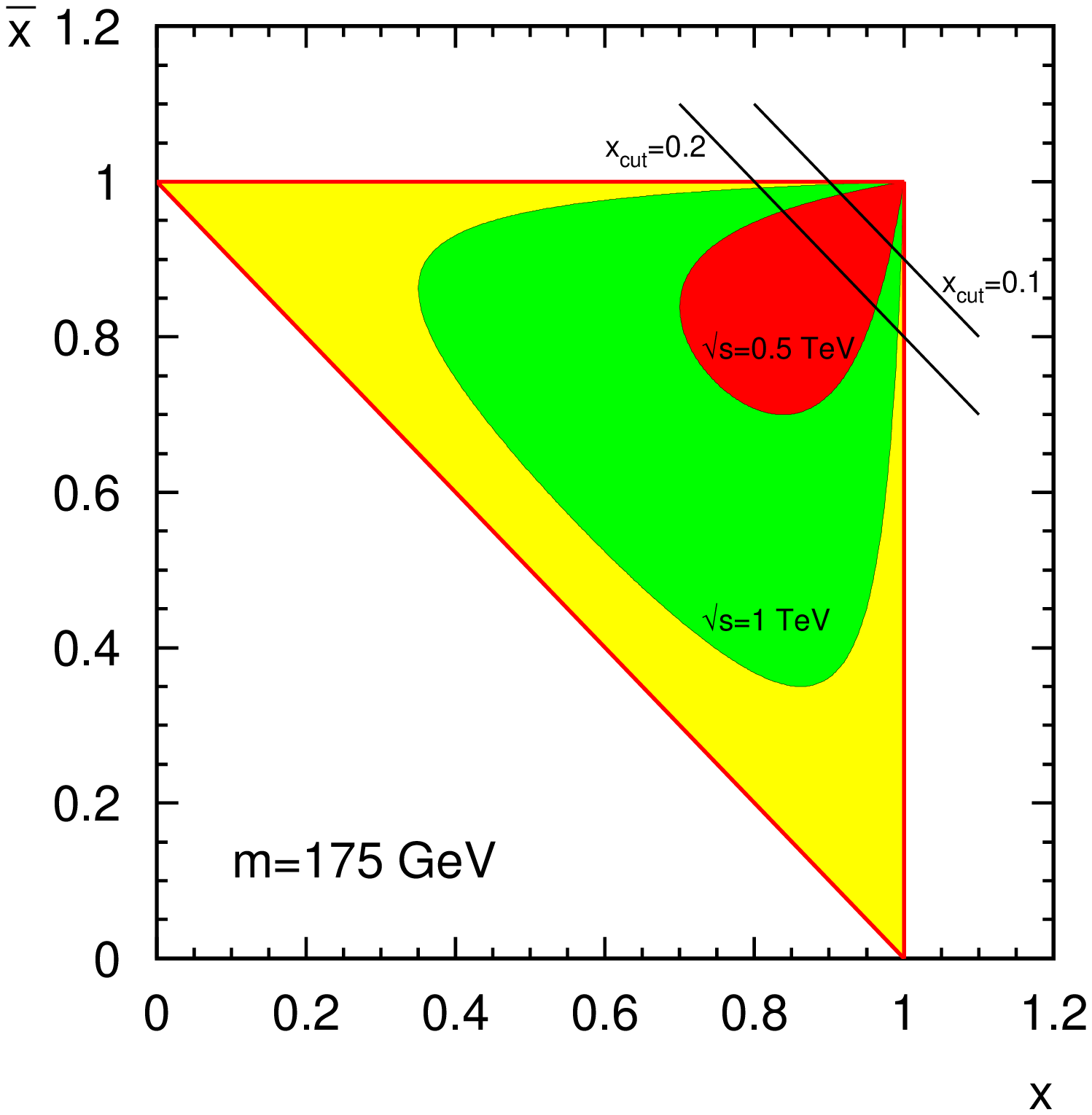,width=9cm,height=9cm}}
\end{picture}
\vskip 0.5cm
\caption{}\label{fig:dalitz}
\end{center}
\end{figure}

\begin{section}{Introduction}
\label{intro}
A future high-energy $e^+e^-$ collider will open up a new domain
for the experimental investigation of the fundamental
interactions between elementary particles.
In particular, the production of top quark pairs in a clean
environment will offer many interesting opportunities both to
perform precision tests of Quantum Chromodynamics (QCD)
and to probe possible deviations from the standard theory.  
For example, the total cross section
for $e^+e^-\to t\bar{t}$ has been computed recently to
next-to-next-to-leading order in the strong coupling 
$\alpha_s$ both in the
threshold region \cite{HoTe98,MeYa98,KuTe99} 
and also far above threshold 
\cite{Che98}. The results obtained for the threshold region   
will allow for a very precise determination of the top quark
mass \cite{BeSiSm99,NaOtSu99,HoTe99}. 

The subject of this paper is the production of top quark pairs
above threshold accompanied by (at least) one additional parton
with a hard momentum. To be more explicit, 
we are interested in the reaction
\bea
\label{reac}
e^+(p_+) + e^-(p_-) \to t(k_t) + \bar{t}(k_{\bar{t}}) + X(k_X),
\eea
and we require that the parton(s)
$X$ carry a large momentum $k_X$. At order $\alpha_s$, the additional
parton can only be a gluon, and at order $\alpha_s^2$ we have the
possibilities $X=g,gg,q\bar{q}$. 
By determining the properties of a final state 
with three or more partons one can 
test the strong interactions in great detail 
as is well known from the
extensive studies of jets in $e^+e^-$ annihilation. For example,
by comparing the cross section for (\ref{reac})
(which will be precisely defined in the next section) 
to the inclusive $t\bar{t}$ production cross section, one
could measure $\alpha_s$. Any deviation from the standard 
determination of $\alpha_s$ using event samples containing 
light quark jets would indicate a violation of the ``flavour
independence'' of the strong interactions --- i.e., would
point towards new physics phenomena.
A specific example of non-standard interactions that could
be probed by reaction (\ref{reac}) are possible anomalous
couplings of the top quark to photons, $Z$-bosons and gluons.
In \cite{Ri96} it has been shown that a 
large anomalous chromomagnetic $t\bar{t}g$ coupling would 
modify the gluon energy spectrum in $e^+e^-\to t\bar{t}g$.
Furthermore, symmetry tests can be performed utilizing 
the richer kinematic structure of the final state in
(\ref{reac}). These contain tests of the CP symmetry 
\cite{BaAtEiSo96} and
the search for final state rescattering effects 
using triple momentum correlations \cite{BrDiSh96}.  
Both the search for heavy quark anomalous
couplings \cite{sld1} and the symmetry tests \cite{sld2} 
have been shown to be experimentally
feasible in the case of $b\bar{b}g$ production at the $Z$
resonance, and it will be interesting to see whether similar
studies are possible with top quarks. 

For any of the above studies it is mandatory to analyse
reaction (\ref{reac}) at next-to-leading order in $\alpha_s$,
which is the topic of this paper. It is organized as follows:  
We start in section \ref{lo} by studying
reaction (\ref{reac}) at leading order. We derive an analytic
formula for the fraction $r$ of $t\bar{t}g$ events with respect to
all $t\bar{t}$ events as a function of the minimal gluon energy.
In section \ref{nlo} we discuss the QCD corrections to reaction
(\ref{reac}).   
We evaluate the fraction $r$ at next-to-leading order for
center-of-mass energies of 0.5 and 1 TeV. We further compute
the differential distribution of the cross section for reaction (\ref{reac}),
normalized to the total $t\bar{t}$ cross section, with respect
to the top quark energy, the $t\bar{t}$ invariant mass, and the
cosine of the angle between the $t$ and $\bar{t}$.
In section \ref{mass} we study the dependence of our results
on both the renormalization scale and the renormalization
scheme employed to define the top quark mass. 
We first review the dependence of the total $t\bar{t}$ cross
section on the mass renormalization scheme by expressing the result
to order $\alpha_s$ in terms of either the perturbative pole mass
or the $\msbar$ (running) mass. We then study in the same fashion 
the scheme and scale dependence of the fraction $r$. 
Finally we estimate, for a c.m.
energy of 1 TeV, the sensitivity of the fraction $r$ on the 
value of the running top quark mass.

\end{section}
\begin{section}{Analysis at leading order}
\label{lo}
In this section we discuss in some detail  
reaction (\ref{reac}) at leading order (LO) in 
$\alpha_s$, i.e., the production of a $t\bar{t}$ 
pair together with a single gluon,
\bea
\label{reaclo}
e^+(p_+) + e^-(p_-) \to t(k_t) + \bar{t}(k_{\bar{t}}) + g(k_g).
\eea
We start by defining the following dimensionless variables:
\bea
\label{scaledvar}
x&=&\frac{2kk_t}{s}=\frac{2E_t}{\sqrt{s}},\nn \\
\xb&=&\frac{2kk_{\bar{t}}}{s}=\frac{2E_{\bar{t}}}{\sqrt{s}},\nn \\
x_g&=&\frac{2kk_g}{s}=\frac{2E_g}{\sqrt{s}}=2-x-\xb,\nn \\
x_{t\bar{t}}&=&\frac{(k_t+k_{\bar{t}})^2}{s}=1-x_g,
\eea
where $\sqrt{s}$ is the center-of-mass energy, 
$k=p_++p_-$, and $E_{t,\bar{t},g}$ are
the energies of the final state particles in the c.m. system. 
The cross section for reaction (\ref{reaclo}) develops 
a soft singularity as $E_g\to 0$. 
An infrared finite cross section may be defined by
demanding 
\bea
\label{cut}
x_g>\xgc,
\eea 
where $\xgc$ is some preset number. 
This condition avoids the region of phase space where the 
gluon becomes soft. Due to the finite mass of the top quark no 
collinear (mass) singularities arise and thus the above condition
is sufficient to define an infrared finite cross section.
The requirement (\ref{cut}) does
{\it not} lead to a finite cross section for $e^+e^-\to q\bar{q}g$,
where $q$ is a massless quark, because the latter cross section
is also singular when the gluon is collinear to the (anti-)quark.
 
We want to study the fraction $r$ of $t\bar{t}g$ events for which 
$x_g>\xgc$ with respect to all $t\bar{t}$ events. 
Since $x_g$ is no useful variable for final states with
four or more partons (which will be relevant
at higher orders in $\alpha_s$), we use instead 
the scaled $t\bar{t}$ invariant mass square $x_{t\bar{t}}$
defined in (\ref{scaledvar}), i.e. replace the condition
$x_g>\xgc$ by the (at LO equivalent) condition:
\bea
\label{xttcut}
1-x_{t\bar{t}}>\xgc.
\eea
The fraction $r$ is defined as
\bea
\label{rlo}
r(\xgc) = \frac{\sigma\left(e^+e^-\to t\bar{t}X;\ 
1-x_{t\bar{t}}>\xgc\right)}
{\sigma_{\rm tot}(e^+e^-\to t\bar{t})}\equiv \frac{\sigma_3(\xgc)}
{\sigma_{\rm tot}}.
\eea
At LO we have $X=g$ and write:
\bea
r(\xgc) =\frac{\alpha_s}{2\pi}\frac{\sigma_3^0(\xgc)}
{\sigma^0_{\rm tot}}
\equiv \frac{\alpha_s}{2\pi}A(\xgc)+O\left(\alpha_s^2\right).
\eea
Here, $\alpha_s/(2\pi)\sigma_3^0(\xgc)$ denotes the 
LO cross section for $e^+e^-\to t\bar{t}g$ with 
$1-x_{t\bar{t}}>\xgc$ and $\sigma^0_{\rm tot}$ is the LO
inclusive cross section for $e^+e^-\to t\bar{t}$.
The coefficient $A(\xgc)$ may be written as follows:
\bea
\label{acoeff}
A(\xgc)=\frac{2N_C\,C_F\sigma_{\rm pt}}{\sigma_{\rm tot}^0}
\int_0^1dx\int_0^1d\xb \,F_1(x,\xb)\Theta(1-x_{t\bar{t}}-\xgc)
\Theta(1-\cos^2\theta_{t\bar{t}}), 
\eea
where $C_F=(N_C^2-1)/(2N_C)$ and $N_C=3$ is the number of colours. 
The point cross section $\sigma_{\rm pt}$ 
reads
\bea 
\label{ptcrosssection}
 \sigma_{\rm pt}\ =\ \sigma(e^+e^- \to \gamma^* \to \mu^+\mu^-)
  \ =\ \frac{4\pi\alpha^2}{3s}, 
\eea
and the LO total cross section $\sigma_{\rm tot}^0$ can be 
expressed in terms of the dimensionless mass variable, 
\bea
\label{zdef}
z = \frac{m_t^2}{s}, 
\eea
as follows:
\bea
\label{sigtotLO}
\sigma_{\rm tot}^0 = N_C\sigma_{\rm pt}\sqrt{1-4z}
\left[c_{V}(1+2z)+ c_{A}(1-4z)\right].
\eea
The coupling factors $c_{V,A}$ appearing in 
(\ref{sigtotLO}) are given explicitly by:
\bea
    \label{wcouplings}
    c_{V} &=&Q_t^2\, f^{\gamma\gamma} 
    + 2 \,g_V^t\,Q_t\, \Re \chi(s)\,f^{\gamma Z} 
    + g_V^{t\,2}\, |\chi(s)|^2\,f^{ZZ},\nonumber\\
    c_{A}  &=& g_A^{t\,2} |\chi(s)|^2 f^{ZZ},
\eea
with
\bea
f^{\gamma\gamma}&=&1-\lambda_-\lambda_+,\nonumber \\
f^{ZZ}&=&(1-\lambda_-\lambda_+)(g_V^{e2}+g_A^{e2})-
      2(\lambda_--\lambda_+) g_V^{e} g_A^{e},\nonumber \\
f^{\gamma Z}&=&-(1-\lambda_-\lambda_+)g_V^e + 
      (\lambda_--\lambda_+)g_A^e.
\eea
Here, $Q_t=2/3$ is the electric charge of the top quark, and
$g_{A,V}^{f}$ denote  the axial and vector couplings of the fermion 
$f$. In particular, $g_V^e = -\frac{1}{2} + 2 \sin^2\vartheta_W$, 
  $g_A^e =-\frac{1}{2}$ for an electron, and  
  $g_V^t = \frac{1}{2} - \frac{4}{3} \sin^2\vartheta_W$,
  $g_A^t = \frac{1}{2}$ for a top quark, where $\vartheta_W$ is the 
  weak mixing angle. The function
  $\chi(s)$ reads
  \begin{equation}
    \label{chi}
    \chi(s) = \frac{1}{4\sin^2\vartheta_W\cos^2\vartheta_W}\,
    \frac{s}{s-m_Z^2 + i m_Z \Gamma_Z},
  \end{equation}
  where $m_Z$ and $\Gamma_Z$ stand for the mass and 
  the width of the Z boson. 
  Finally,
  $\lambda_-$ ($\lambda_+$) denotes the longitudinal 
  polarization  of the electron (positron) beam.
\par
The function $F_1$ may be decomposed as:
\bea
\label{Fi01}
F_1(x,\xb) &=& c_{V}F_1^{V}(x,\xb)+c_{A}F_1^{A}(x,\xb).
\eea 
Using the abbreviation
\bea 
B =\frac{1}{(1-x)(1-\xb)},
\eea
we have \cite{BrUw98}: 
\bea
\label{Fi02}
F_1^{V}(x,\xb) &=& B\bigg\{ {x^2+\xb^2 \over 2 }
  -  z \left[2x_g+\left((1-x)^2+(1-\xb)^2\right)B \right]
  - 2z^2x_g^2B\bigg\}, 
\eea
\bea
F_1^{A}(x,\xb) &=& F_1^{VV}(x,\xb) + z
B\bigg\{(x+\xb)^2-10(1-x_g)+6zx_g^2B\bigg\}. 
\eea
\par
The cosine of the angle between the top quark and antiquark appearing
in (\ref{acoeff}) can be 
expressed in terms of the top quark mass and the scaled energy variables:
\bea
\label{costheta}
\cos\theta_{t\bar{t}}=\frac{x\xb-2(1-x_g)+4z}
{\sqrt{(x^2-4z)(\xb^2-4z)}}.
\eea
The Heaviside function $\Theta(1-\cos^2\theta_{t\bar{t}})$ defines
the kinematically allowed region in the $(x,\xb)$ plane
when no cuts are applied. 
For $m_t=175$ GeV and the two c.m. energies $\sqrt{s} = 0.5$ TeV 
and  $\sqrt{s}= 1$ TeV this region is depicted in
Fig. \ref{fig:dalitz}. The cuts on the scaled gluon
energy $\xgc=0.1,0.2$ are indicated as lines, and the enveloping
triangle is the kinematically allowed region for massless quarks. 
With a cut $\xgc$ on the gluon energy, the kinematically allowed
area $P(\xgc,z)$ is:
\bea
P(\xgc,z) &\equiv& \int_0^1 dx \int_0^1 d\xb\,
\Theta(1-x_{t\bar{t}}-\xgc)
\Theta(1-\cos^2\theta_{t\bar{t}})\nn \\ 
&=& \Theta(1-4z-\xgc)\left[2z(1-z)\ln\left(\rho\right)
+\frac{1}{2}(1-\xgc)(1+\xgc+2z)\frac{1-\rho}{1+\rho}\right],
\eea
where
\bea
\rho &=& \frac{1-\sqrt{1-4z/(1-\xgc)}}
{1+\sqrt{1-4z/(1-\xgc)}}.
\eea
When the kinematic boundary $1-4z=\xgc$ is approached for
a fixed value of $\xgc$, $P(\xgc,z)$ 
vanishes like  $(1-4z-\xgc)^{3/2}$.
\par
The function $A(\xgc)$ defined in (\ref{acoeff}) can be
very easily computed numerically. This has the advantage 
that an implementation of additional and/or different cuts is
straightforward. On the other hand, an analytic result
for $A(\xgc)$ is obviously desirable, and in fact it is possible
to perform the twofold integral in (\ref{acoeff}) analytically.
The result is:
\bea
\label{ares1}
A(\xgc)=\frac{2N_C\,C_F\sigma_{\rm pt}}{\sigma_{\rm tot}^0} 
 \Theta(1-4z-\xgc)\left[c_V A^{V}(\xgc)+c_A A^{A}(\xgc)\right],
\eea 
where
\bea
\label{ares2}
A^{V}(\xgc)&=&
-\frac{1}{2}(1+4\xgc-\xgc^2+8z\xgc+2z^2)\ln(\rho)
\nn \\ &+& \frac{1}{4}(1-\xgc)(11-3\xgc+34z)\frac{1-\rho}{1+\rho}  -
(1-4z^2)g(\omega,\rho),\nn \\
A^{A}(\xgc)&=&A^{V}(\xgc)-
z\Big[(1-4z-12\xgc-\xgc^2+6z^2)\ln(\rho) \nn \\ 
&+& \frac{1}{2}(1-\xgc)(51+\xgc-6z)\frac{1-\rho}{1+\rho} - 
6(1-2z)g(\omega,\rho)\Big],
\eea
with
\bea
g(\omega,\rho)&=&\ln^2(\omega)-2\ln\left(\frac{\rho-\omega}
{1-\rho\,\omega}\right)\left(\ln(\omega)+\frac{1-\omega^2}
{1+\omega^2}\right)
\nn \\ &+&\ln^2(\rho)+
2\,\Li{1-\frac{\rho}{\omega}}+2\,\Li{1-\rho\,\omega}
\eea
and
\bea
\omega &=& \frac{1-\sqrt{1-4z}}{1+\sqrt{1-4z}}.
\eea
\par
Fig. \ref{fig:ecm} shows the fraction $r$ defined in (\ref{rlo})
as a function of the c.m. energy 
for $m_t=175$ GeV, and three different values of $\xgc$.
Here and in all other numerical results of this paper 
we consider unpolarized electron and positron beams.
The running of the strong coupling is taken into 
account in the curves,
i.e. we use $\alpha_s(\mu=\sqrt{s})$ with six active flavours.
Our input value is $\alpha_s^{n_f=5}(\mu=m_Z)=0.118$, which is 
evolved up
to $\mu=m_t$ with five active flavours, and then converted
into $\alpha_s^{n_f=6}(m_t)$ using 
the so-called matching conditions of 
\cite{BeWe80}{\footnote{To be more precise,
the conversion is performed at the scale
$\overline{m}_t(\overline{m}_t)$, where $\overline{m}_t$ is the 
running top quark mass.}}.
For example, we thus get $\alpha_s^{n_f=6}(\mu=500 {\mbox{ GeV}})=0.0952$.
\par
We close this section with a remark concerning the experimental
distinction of $t\bar{t}$ events with a gluon radiated off the $t$
or $\bar{t}$ from events in which the gluon is radiated off
the $b$ or $\bar{b}$ produced in the decays of the top quark pairs.
It has been shown in \cite{MaOr98} that the following two constraints
efficiently select events where the gluon is 
produced in association with 
the top quark pair:
\bea
\label{orr}
E_g &>& \frac{\sqrt{s}}{2}\xgc \gg \Gamma_t,  \nn \\
m_t-2\Gamma_t &\le& \sqrt{\left(k_{W^{\pm}}+k_{b(\bar{b})}\right)^2}
\le m_t+2\Gamma_t.
\eea
By requiring that the invariant mass of the $Wb$ system lies in the
vicinity of the top quark mass, the probability that a highly
energetic gluon jet  
($E_g \gg \Gamma_t \approx 1.4$ GeV) 
is emitted from the $b$ or $\bar{b}$ is very small. Our computation
may therefore be applied to describe events of the type
$e^+e^-\to W^+W^-b\bar{b}g$  that fulfil both conditions of 
(\ref{orr}). 
Our ``default''
value for many of the results below will be $\xgc=0.1$, which
corresponds to $E_g=25\ (50)$ GeV for a c.m. energy
$\sqrt{s}=0.5\ (1)$ TeV.

\end{section}
\begin{section}{Results at next-to-leading order}
\label{nlo}
At order $\alpha_s^2$, we have to consider both virtual and
real corrections to the process (\ref{reaclo}). 
The real corrections consist of the processes
$e^+e^- \to t\bar{t}gg,\ e^+e^- \to t\bar{t}q\bar{q}\ 
(q=u,d,s,c,b)$. If $\sqrt{s}>4m_t$, the production of
{\it two} $t\bar{t}$ pairs becomes possible. However, these
rather spectacular events are extremely rare for the c.m. energies
considered below and contributions from 
the process $e^+e^-\to t\bar{t}t\bar{t}$
can therefore be neglected{\footnote{Contributions from
``secondary'' $t$-quarks, originating from gluon 
splitting in $e^+e^-\to q\bar{q}g$, are also heavily suppressed.}}.
\par    
As mentioned before, the condition $1-x_{t\bar{t}}>\xgc$ 
is equivalent, at LO, to 
requiring a minimal scaled gluon energy $x_g$. 
For final states with four or more partons, 
an alternative condition 
is $2kk_X/s>\xgc$, where  in the c.m.
system $2kk_X/s$ is equal to $2E_X/\sqrt{s}$, with 
$E_X= \sqrt{s}- E_t- E_{\bar{t}}$. 
For a given cut $\xgc$, this condition leads to
larger contributions to $r(\xgc)$ from the four-parton final states
as compared to the definition
employed in (\ref{rlo}), since
$2E_X/\sqrt{s}=1-x_{t\bar{t}}+k_X^2/s$ with $k_X^2\ge 0$. 

We renormalize the coupling in the modified minimal subtraction
($\msbar$) scheme. For the results of this section, the
top quark mass is defined as the perturbative pole mass, i.e., 
the mass renormalization is carried out in the on-shell scheme.
The result for the fraction $r(\xgc)$ 
at a renormalization scale $\mu$ may be written to 
next-to-leading accuracy as
\bea
\label{rnlo2}
r(\xgc,\mu) =
\frac{\sigma_3(\xgc,\mu)}
{\sigma_{\rm tot}(\mu)}= 
\frac{\alpha_s(\mu)}{2\pi}A(\xgc)
+\left(\frac{\alpha_s(\mu)}{2\pi}\right)^2B(\xgc,\mu)+O\left(\alpha_s^3\right),
\eea
with
\bea
\sigma_3(\xgc,\mu) = \frac{\alpha_s(\mu)}{2\pi}
\sigma_3^0(\xgc)
+\left(\frac{\alpha_s(\mu)}{2\pi}\right)^2
\sigma_3^1(\xgc,\mu)+O\left(\alpha_s^3\right),
\eea
\bea
\sigma_{\rm tot}(\mu)=\sigma_{\rm tot}^0
+\frac{\alpha_s(\mu)}{2\pi}\sigma_{\rm tot}^1+O\left(\alpha_s^2\right).
\eea
The coefficient $A(\xgc)$ is given explicitly in Eqs. 
(\ref{ares1}), (\ref{ares2}),  and the coefficient
$B(\xgc,\mu)$ reads:
\bea
B(\xgc,\mu) = 
\frac{\sigma_3^1(\xgc,\mu)}
{\sigma_{\rm tot}^0} - 
\frac{\sigma_{\rm tot}^1\sigma_3^0(\xgc)}
{\left(\sigma_{\rm tot}^0\right)^2}.
\eea
The result for $\sigma_{\rm tot}^1$ is well
known (see, e.g. \cite{ChKuKw96}).
For the computation of $\sigma_3^1(\xgc,\mu)$ 
we use the techniques and  
results of \cite{BeBrUw97} and \cite{BrUw98}. In these papers,
the production of three jets involving $b$ quarks
was studied at NLO, taking into account the $b$ quark mass. 
The matrix elements given there can be easily adapted
to the case at hand. For technical details of the 
calculation, we refer the reader to
\cite{BrUw98}. 

Figure 3 (4) shows the LO and NLO results
for $r$ as a function of $\xgc$ at $\sqrt{s}=0.5$ TeV 
($\sqrt{s}=1$ TeV). The renormalization scale is set to
$\mu=\sqrt{s}$. (The scale dependence of our
results will be discussed in the next section.) 
In Table 1 we list the $A$ and $B$ 
coefficients for a sample of $\xgc$-values. 
The numerical errors in the last digits of the $B$ 
coefficients are also shown{\footnote{For example, $57.1(10)$ stands
for $57.1\pm 1.0$.}}.
For $\sqrt{s}=0.5$ TeV, the relative 
size of the QCD corrections, $\alpha_s/(2\pi)B/A$,
varies between 56\% (at $\xgc=0.02$) and 36\% (at $\xgc=0.2$).
At $\sqrt{s}=1$ TeV, the QCD corrections are roughly
constant as $\xgc$ is varied and of the order of 30\%.
In Figs. 5, 6, and 7 we plot various distributions of the 
cross section $\sigma_3(\xgc,\mu)$. All distributions are
normalized to the total $t\bar{t}$ cross section $\sigma_{\rm tot}$,
and we set $\sqrt{s}=\mu=0.5$ TeV, $m_t=175$ GeV, and 
$\xgc=0.1$. These values lead to the kinematic limits  
$0.7\le x \lesssim 0.9837$, 
$0.1\le 1-x_{t\bar{t}}\le 0.51$ for the scaled top quark energy
$x$ and the quantity $1-x_{t\bar{t}}$, respectively.
(Recall that in leading order $1-x_{t\bar{t}}$
is equal to the scaled gluon energy $x_g$.)  
Fig. 5 shows the distribution $1/\sigma_{\rm tot}d\sigma_3/dx$.
The distribution reaches 
its maximum around $x\approx 0.925$,
which corresponds to $E_t \approx 231$ GeV. 
We note in passing that the distribution with respect to the scaled
top antiquark energy $\xb$ is equal to the one shown due
to charge conjugation invariance of the strong interactions.
In Fig. 6 we plot the distribution 
$1/\sigma_{\rm tot}d\sigma_3/d(1-x_{t\bar{t}})$.
The size of the QCD corrections depends
only weakly on $1-x_{t\bar{t}}$. (The seemingly large 
correction in the last bin is most probably a fluctuation
due to the limited statistics of the Monte Carlo integration.)
Fig. 7 depicts the distribution with respect to the cosine
of the angle between the top quark and antiquark. As one might
have expected, this distribution is very sharply peaked
close to  $\cos\theta_{t\bar{t}}=-1$.   
\end{section}

\begin{section}{Top quark pole mass versus running mass}
\label{mass}
All results in the preceding section have been obtained by using the
$\msbar$ scheme for the coupling renormalization and by 
renormalizing the top quark mass in the on-shell scheme.
In this section we study the dependence of our results
on the mass renormalization scheme as well as on the
choice of the renormalization scale $\mu$. 

In order to understand the impact of a change of the  
mass definitions on our fixed order predictions, it is
instructive first to consider the simple and well-known 
case of the inclusive rate $\sigma_{\rm tot}$.
We use the notation $z^{\rm on}=\left(m_t^{\rm on}\right)^2/s$ and
$\overline{z}(\mu)=\overline{m}_t^2(\mu)/s$, where $m_t^{\rm on}$ is the
pole mass (for which we always use the value $m_t^{\rm on}=175$ GeV) 
and $\overline{m}_t(\mu)$ is the $\msbar$ mass at a scale $\mu$.
The latter is obtained from the renormalization group evolution with
the input $\overline{m}_t(\mu=\overline{m}_t)=168$ GeV.
The relation between the two mass parameters 
is given by 
\bea
z^{\rm on} = \overline{z}(\mu)+\Delta\overline{z}(\mu),
\eea
where, to order $\alpha_s$,  
\bea
\Delta\overline{z}(\mu) = \frac{2C_F\alpha_s(\mu)}{\pi}
\left[1-\frac{3}{4}\ln\left(\frac{\overline{z}(\mu)s}{\mu^2}\right)\right]
\overline{z}(\mu)
+O\left(\alpha_s^2\right).
\eea
To order $\alpha_s$ we thus have{\footnote{I thank M. Spira for explaining
to me this method of switching between two mass renormalization schemes.}}:
\bea
\label{sigconv}
\sigma_{\rm tot}(z^{\rm on},\mu)
&=&\sigma_{\rm tot}(\overline{z}(\mu)+\Delta\overline{z}(\mu),\mu)\nn \\
&=&\sigma_{\rm tot}^0(\overline{z}(\mu)) + 
\Delta\overline{z}(\mu)
\frac{d\sigma_{\rm tot}^0}{d z}\left(\overline{z}(\mu)\right)
+\frac{\alpha_s(\mu)}{2\pi}\sigma_{\rm tot}^1(\overline{z}(\mu))
+O\left(\alpha_s^2\right),
\eea 
where  $d\sigma_{\rm tot}^0/d z$ can be obtained from
(\ref{sigtotLO}). 
Fig. 8 (9) shows the total cross section in units of the point cross section
(\ref{ptcrosssection})
as a function of the renormalization scale at $\sqrt{s}=0.5$ TeV 
($\sqrt{s}=1$ TeV) for both mass definitions. When $\sigma_{\rm tot}$
is expressed in terms of the pole mass,
the dependence on $\mu$ is (to order $\alpha_s$) 
only induced by the running of $\alpha_s$.
Two important features are visible in the curves:
First, within a large range of values for $\mu$, the QCD corrections
are smaller when the result is expressed in terms of the running 
mass.
Second, the $\mu$ dependence is flatter if the running mass is used. 
For small values of $\mu$ one further observes a strong decrease 
(increase) of the LO (NLO) prediction for
$\sigma_{\rm tot}$ expressed in terms of the running mass. 
These unphysical
features occur because {\it both} the coupling {\it and} 
the running mass become large as $\mu$ becomes small{\footnote{
For example, at $\mu=25$ GeV we have 
$\overline{m}_t(\mu=25{\mbox{ GeV}})\approx 199.3$ GeV.}}.
 The decrease of the 
LO result simply reflects the fact that 
$\sigma^0_{\rm tot}$ goes to zero
as $z$ approaches the threshold value $z=1/4$. The fact 
that the NLO result in both mass renormalization
schemes is rather stable
when $\mu$ is varied as well as the good agreement of the predictions
in the two schemes indicates that the contributions of order $\alpha_s^2$ are
small; this is indeed the case, as the explicit calculation of \cite{Che98}
shows. 
  
We now turn back to the discussion of the fraction $r(\xgc)$. 
The renormalization scale dependence of the NLO coefficient $B$ 
in the pole mass scheme is easily obtained:
\bea
B(\xgc,z^{\rm on},\mu)=B(\xgc,z^{\rm on},\sqrt{s})
-A(\xgc,z^{\rm on})\beta_0
\ln\left(\frac{\sqrt{s}}{\mu}\right),
\eea
with $\beta_0=(11N_C-2n_f)/3$ and $n_f=6$. 
In analogy to
(\ref{sigconv}) we may express the result for $r(\xgc)$
in terms of the $\msbar$ mass at the scale $\mu$ by writing
\bea
r(\xgc,z^{\rm on},\mu) &=& 
r(\xgc,\overline{z}(\mu)+\Delta\overline{z}(\mu),\mu)\nn \\
&=& \frac{\alpha_s(\mu)}{2\pi}A\left(\overline{z}(\mu)\right)
+\Delta\overline{z}(\mu)\frac{d A}{d z}\left(\overline{z}(\mu)\right)
\nn \\
&+&\left(\frac{\alpha_s(\mu)}{2\pi}\right)^2B(\xgc,\overline{z}(\mu),\mu)+
O\left(\alpha_s^3\right).
\eea 
Figs. 10 and 11 show the LO and NLO results for $r$ at $\xgc=0.1$
as a function of $\mu$ both in the pole mass and the running mass
scheme at $\sqrt{s}=0.5$ TeV and $\sqrt{s}=1$ TeV, respectively.
At $\sqrt{s}=0.5$ TeV, we find
a rather large difference of the NLO results in the
two schemes at scales $\mu \sim \sqrt{s}$. 
The additional gluon 
in the final state carries (at $\xgc=0.1$) a momentum
of at least 25 GeV, and the allowed phase space depends rather
strongly on the top quark mass{\footnote{The running
top quark mass decreases from $199.3$ GeV to $155.3$ GeV as 
$\mu$ is varied between 25 and 500 GeV.}}, which explains the strong 
$\mu$-dependence of the LO result in terms of the running mass.
At NLO the $\mu$-dependence is drastically reduced, but 
$r$ still increases with $\mu$. The gain in phase space
due to the decrease of $\overline{m}_t(\mu)$ 
overcompensates the decrease of $\alpha_s(\mu)$ even at NLO.
This suggests  
that the running mass should not be used in the
fixed-order prediction for the fraction $r$ 
at $\sqrt{s}=0.5$ TeV and $\xgc\ge 0.1$. 
At these values,
we are too close to the kinematic threshold of the $t\bar{t}g$
final state where the running mass is an ``unnatural'' parameter. 

At $\sqrt{s}=1$ TeV the differences of the
NLO results for the two mass definitions are smaller as compared
to the case of $\sqrt{s}=0.5$ TeV. Also, the NLO results only
weakly depend on the choice of $\mu$, especially if the
running mass is used. The two scales $m_t$ and $\sqrt{s}$
are now rather far apart, and the running mass parameter 
becomes preferable. For even higher c.m. energies, the fraction
$r$ as defined in (\ref{rlo}) eventually suffers from large
contributions in the collinear region; as $z\to 0$, the  result
in $n$th order
is dominated by terms $\sim \left(\alpha_s\ln(z)\right)^n$, 
which should either be resummed or avoided by modifying the
definition of $r$. One could for example,
after selecting samples of  $t\bar{t}X$ events with 
$1-x_{t\bar{t}}>\xgc$, use in addition standard 
jet algorithms to define collinear safe cross sections.

An intriguing question is whether a measurement of
the fraction $r$ allows a direct determination of the
value of the running mass parameter of the top quark at
high energies. (Analogously, NLO results for three-jet fractions
involving $b$ quarks \cite{RoSaBi97,BeBrUw97,BrUw98}
have been used to extract a value for $\overline{m}_b(\mu=m_Z)$
from the high-statistics LEP \cite{delphi98} and SLD 
\cite{bu98,bu99} data.) We consider here as a case study
the result for $r(\xgc=0.1)$ at $\sqrt{s}=1$ TeV. As was discussed
above, for these values the fraction $r$ 
is perturbatively well under control when expressed in terms 
of the running mass. We now vary the value $\overline{m}_t(\mu=
1 {\mbox{ TeV}})$ between 140 and 160 GeV{\footnote{The
value obtained from the renormalization group evolution
is 
$\overline{m}_t(\mu=1{\mbox{ TeV}})=148.6$ GeV.}} and compute 
$r(\xgc=0.1)$ at LO and NLO. The results 
are shown in Fig. 12.
At NLO, $r(\xgc=0.1)$ decreases by about
4\% when the running top quark mass is changed from 140 to 
160 GeV. If a measurement of $r$ will be possible with
an error of $\pm 1$\%, the running top quark mass
at $\mu=1$ TeV could be determined up to $\pm 5$ GeV.
A statistical error of 1\% on $r$ is realistic 
with the envisioned high luminosity of a 
future linear collider operating at $\sqrt{s}=1$ TeV, but whether
or not the systematic error can be kept that small remains 
to be seen.

Although the sensitivity of the fraction $r$ 
on $\overline{m}_t(\mu=1 {\mbox{ TeV}})$
seems to be rather poor, a direct determination of the running
top quark mass at such a high scale would provide a nice
test of perturbative QCD in the following way:
It is expected that a very precise
value of the top quark mass can be obtained 
from the threshold scan of $\sigma_{\rm tot}$. 
Since the pole mass suffers
from (nonperturbative) infrared ambiguities, improved
mass definitions have been proposed in this context,
the so-called potential-subtracted \cite{BeSiSm99,NaOtSu99}
and 1S \cite{HoTe99} mass. After extracting these masses 
from the threshold scan, they can be converted into the
$\msbar$ mass, evolved to $\mu=1$ TeV, and compared to the
direct measurement suggested above. 
 
\end{section}

\begin{section}{Conclusions}
\label{concl}
At a future high-energy and high-luminosity $e^+e^-$ linear
collider, the production of top quark pairs in association
with one or more additional hard partons will be an interesting
new testing ground for perturbative QCD. For example,
a measurement of the fraction $r$ of 
$t\bar{t}X$ events with respect to all $t\bar{t}$ 
events will provide a powerful ``flavour independence
test'' of the strong interactions.
In this paper we have studied the reaction $e^+e^-\to t\bar{t}X$
to order $\alpha_s^2$.
The fraction $r$ has been defined as a function of a 
preset cut parameter $\xgc$ 
by demanding 
$1-x_{t\bar{t}}>\xgc$ where $x_{t\bar{t}}=(k_t+k_{\bar{t}})^2/s$
is the scaled invariant mass square of the $t\bar{t}$ pair. 
At LO, where $1-x_{t\bar{t}}$ is equal to the scaled gluon energy 
$x_g=2E_g/\sqrt{s}$, 
we have calculated this fraction analytically.
The QCD corrections to  $r(\xgc)$, which we have evaluated 
for c.m. energies $\sqrt{s}=0.5$ TeV and $\sqrt{s}=1$ TeV, 
are large and positive. We have further computed distributions
of the differential cross section for $e^+e^-\to t\bar{t}X$ with respect
to the scaled top quark energy, the variable $1-x_{t\bar{t}}$,
and the cosine of the angle between $t$ and $\bar{t}$.
A measurement of such distributions will be interesting 
for instance in the context of searches
for possible anomalous top quark couplings, and the QCD
corrections have to be included in such analyses.
Finally we have studied the dependence 
of our results on the renormalization
scale $\mu$ and compared the pole mass and the running
mass renormalization schemes. We have found that
at $\sqrt{s}=0.5$ TeV the phase space for $t\bar{t}g$ 
production depends (for a hard gluon) 
rather strongly on the value of the
top quark mass parameter. We have concluded that for this
(or smaller) c.m. energies the fixed order prediction for the
fraction $r$ should not be expressed in terms of the running mass.
At $\sqrt{s}=1$ TeV, $\overline{m}_t(\mu)$ becomes the preferable
mass parameter. This is reflected 
in the improved stability of the NLO
results under variations of the scale $\mu$. If a measurement
of $r$ at $\sqrt{s}=1$ TeV 
is possible to an accuracy of 1\%, the running mass
$\overline{m}_t(\mu=1 {\mbox{ TeV}})$
could be directly determined up to about $\pm 5$ GeV, offering
a nice consistency check of the renormalization group
evolution of $\overline{m}_t(\mu)$ as predicted by perturbative 
QCD.

In this paper we have concentrated on studying 
the fraction of $t\bar{t}X$ events and their
distributions with respect to energies of and angles between
final state particles only. Future work will include the
investigation of observables that involve the orientation 
of the final state with respect to the electron beam, 
like forward-backward asymmetries or so-called 
``event handedness'' correlations \cite{BrDiSh96}. 
\end{section}
\section*{Acknowledgements}
I would like to thank W. Bernreuther, M. Flesch, M. Spira, and
P. Uwer for many enlightening discussions.

\newpage
\begin{center}
TABLE CAPTIONS
\end{center}
\noindent 
{\bf Table 1.} Leading order and next-to-leading order coefficients
of the fraction $r(\xgc,\mu)$ as defined in Eq. (\ref{rnlo2}) for two
different center-of-mass energies and $\xgc=0.1$. The renormalization scale
$\mu$ is set to $\sqrt{s}$, and a top quark pole mass 
of $m_t=175$ GeV is used.   
\begin{center}
FIGURE CAPTIONS
\end{center}
\noindent 
{\bf Fig. 1.} Kinematically allowed region in the $(x,\xb)$ plane
for a $t\bar{t}g$ final state with $m_t=175$ GeV. The enveloping
light-coloured triangle is the allowed region for massless quarks.
Lines indicate cuts on the scaled gluon energy.\\
{\bf Fig. 2.} Fraction $r(\xgc)$ (defined in (\ref{rlo})) 
at leading order in $\alpha_s$  for
$\xgc=0.02,0.1,0.2$ and $m_t=175$ GeV as a function of 
the center-of-mass energy
$\sqrt{s}$. The running of $\alpha_s$ is included.\\
{\bf Fig. 3.} Fraction $r(\xgc)$ at leading (LO) and next-to-leading
order (NLO) as a function of $\xgc$ at $\sqrt{s}=0.5$ TeV. 
The renormalization
scale $\mu$ is set to $\sqrt{s}$.\\
{\bf Fig. 4.} Same as Fig. 3, but for $\sqrt{s}=1$ TeV.\\
{\bf Fig. 5.} Distribution $1/\sigma_{\rm tot}d\sigma_3/dx$ at LO
and NLO for $\sqrt{s}=\mu=0.5$ TeV and $\xgc=0.1$. 
The scaled quark energy is defined in Eq. (\ref{scaledvar}),
and the cross section $\sigma_3$ is defined in Eq. (\ref{rnlo2}).\\
{\bf Fig. 6.} Same as Fig. 5, but for  
$1/\sigma_{\rm tot}d\sigma_3/d(1-x_{t\bar{t}})$ with $x_{t\bar{t}}$
defined in Eq. (\ref{scaledvar}).\\
{\bf Fig. 7.} Same as Fig. 5, but for 
$1/\sigma_{\rm tot}d\sigma_3/d(\cos\theta_{t\bar{t}})$ with
$\cos\theta_{t\bar{t}}$ defined in Eq. (\ref{costheta}).\\
{\bf Fig. 8.} LO and NLO results for the total $t\bar{t}$ cross
section in units of the point cross section defined 
in Eq. (\ref{ptcrosssection})
at a  c.m. energy  of $\sqrt{s}=0.5$ TeV. 
The renormalization scale $\mu$ is varied 
between 25 GeV and $\sqrt{s}$, and results obtained in
the pole mass and in the $\msbar$ (running) mass 
renormalization schemes are compared.\\
{\bf Fig. 9.} Same as Fig. 8, but for $\sqrt{s}=1$ TeV.\\
{\bf Fig. 10.} Same as Fig. 8, but for the fraction
$r$ at $\xgc=0.1$.\\
{\bf Fig. 11.} Same as Fig. 10, but for $\sqrt{s}=1$ TeV.\\
{\bf Fig. 12.} Dependence of the fraction $r$ at $\xgc=0.1$
and $\sqrt{s}=\mu=1$ TeV on the value of the running top
quark  mass $\overline{m}_t$ at the scale $\mu=1$ TeV.
The dashed line is the LO result, and the solid line is the 
NLO result.  

\newpage
\par
    \begin{table}
      \begin{center}
        \begin{tabular}[b]{c|c|c|c|c|}
\ &  \multicolumn{2}{c|}{$\sqrt{s}=0.5$ TeV} &  
\multicolumn{2}{c|}{$\sqrt{s}=1$ TeV} \\
\hline
$\xgc$ & $A$ & $B$ & $A$ & $B$ 
 \\
          \hline
0.02 & 10.02 & 371(2) & 35.46 & 680(3)  \\ 
\hline
0.04 & 6.976 & 224(2) & 26.87 & 524(3) \\ 
\hline
0.06 & 5.302 & 156(2) & 21.99 & 447(3) \\ 
\hline
0.08 & 4.185 & 119(2) & 18.64 & 384(3) \\ 
\hline
0.10 & 3.373 & 90.6(12) & 16.12 & 347(3) \\ 
\hline
0.12 & 2.751 & 71.5(10) & 14.13 & 309(3) \\ 
\hline
0.14 & 2.260 & 57.1(10) & 12.50 & 280(3)\\ 
\hline
0.16 & 1.864 & 46.9(10) & 11.12 & 251(2) \\ 
\hline
0.18 & 1.539 & 36.9(10) & 9.957 & 226(2) \\ 
\hline
0.20 & 1.271 & 30.1(8) & 8.945 & 200(2)\\ 
\hline
        \end{tabular}
      \end{center}
      \caption{}\label{tab1}
    \end{table}

\newpage
\ 
\begin{figure}[ht]
\unitlength1.0cm
\begin{center}
\begin{picture}(8,8)
\put(0,0){\psfig{figure=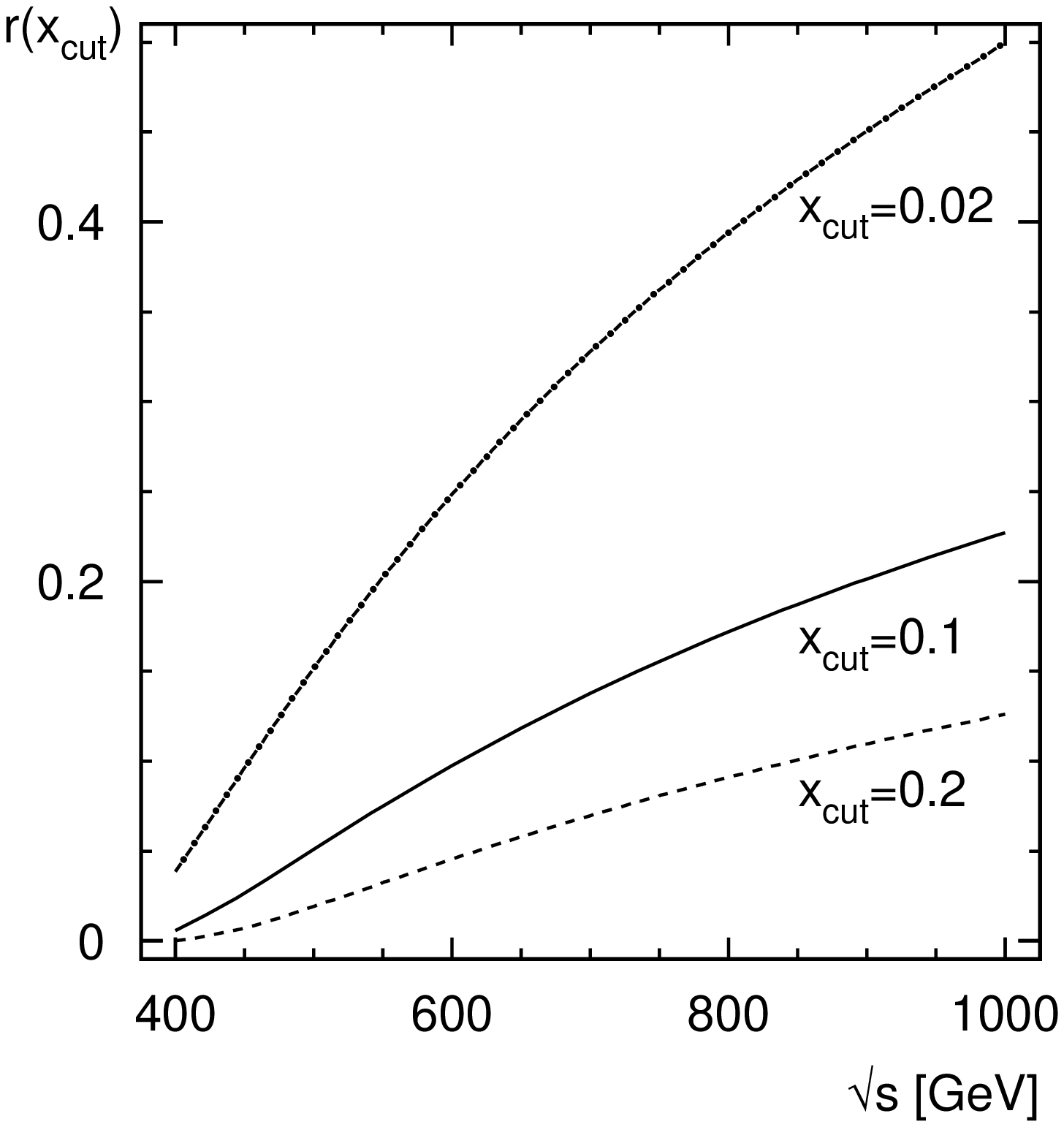,width=9cm,height=9cm}}
\end{picture}
\vskip 0.5cm
\caption{}\label{fig:ecm}
\end{center}
\end{figure}
\begin{figure}[ht]
\unitlength1.0cm
\begin{center}
\begin{picture}(8,8)
\put(0,0){\psfig{figure=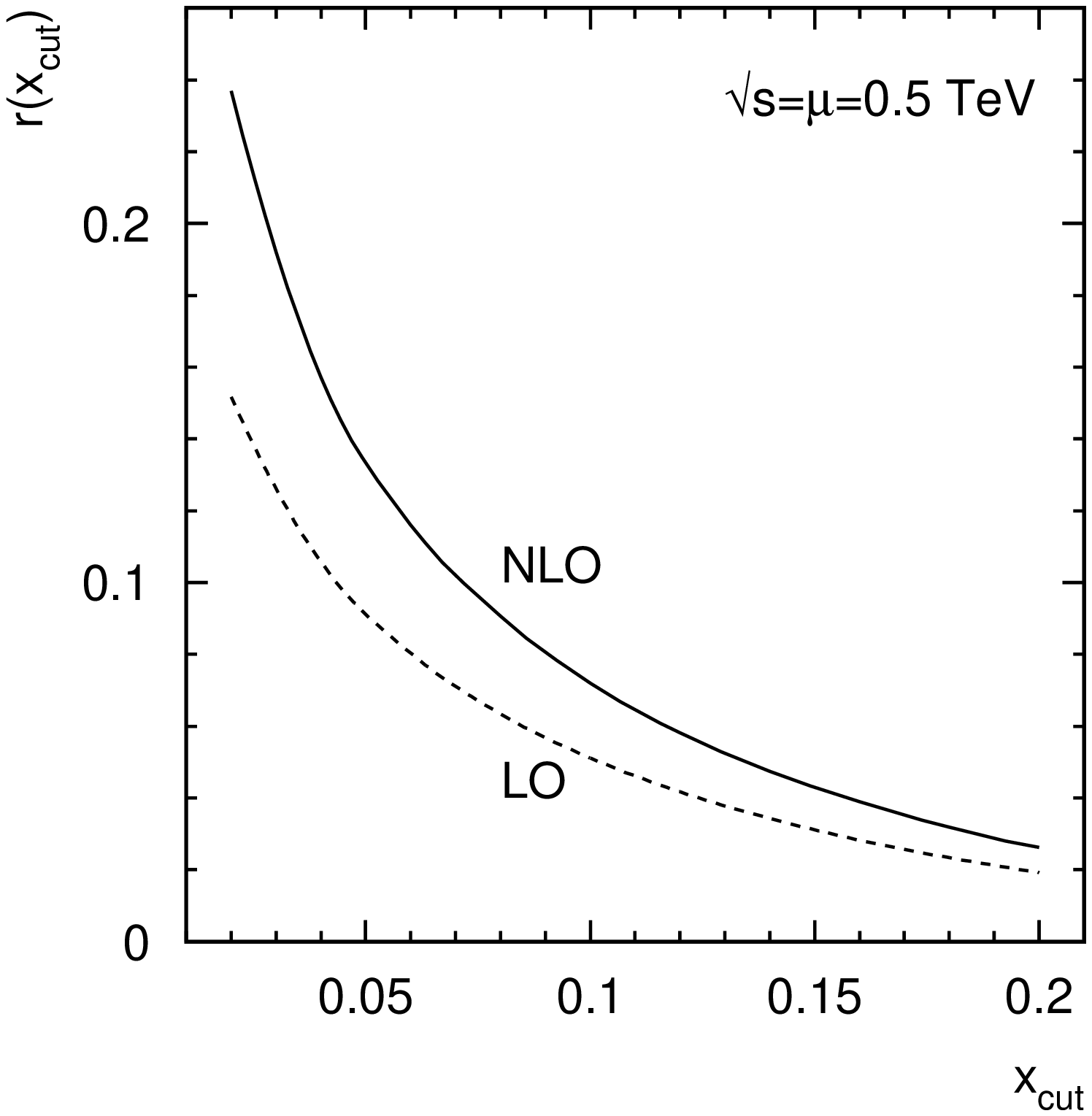,width=9cm,height=9cm}}
\end{picture}
\vskip 0.5cm
\caption{}\label{fig:ycut}
\end{center}
\end{figure}
\begin{figure}[ht]
\unitlength1.0cm
\begin{center}
\begin{picture}(8,8)
\put(0,0){\psfig{figure=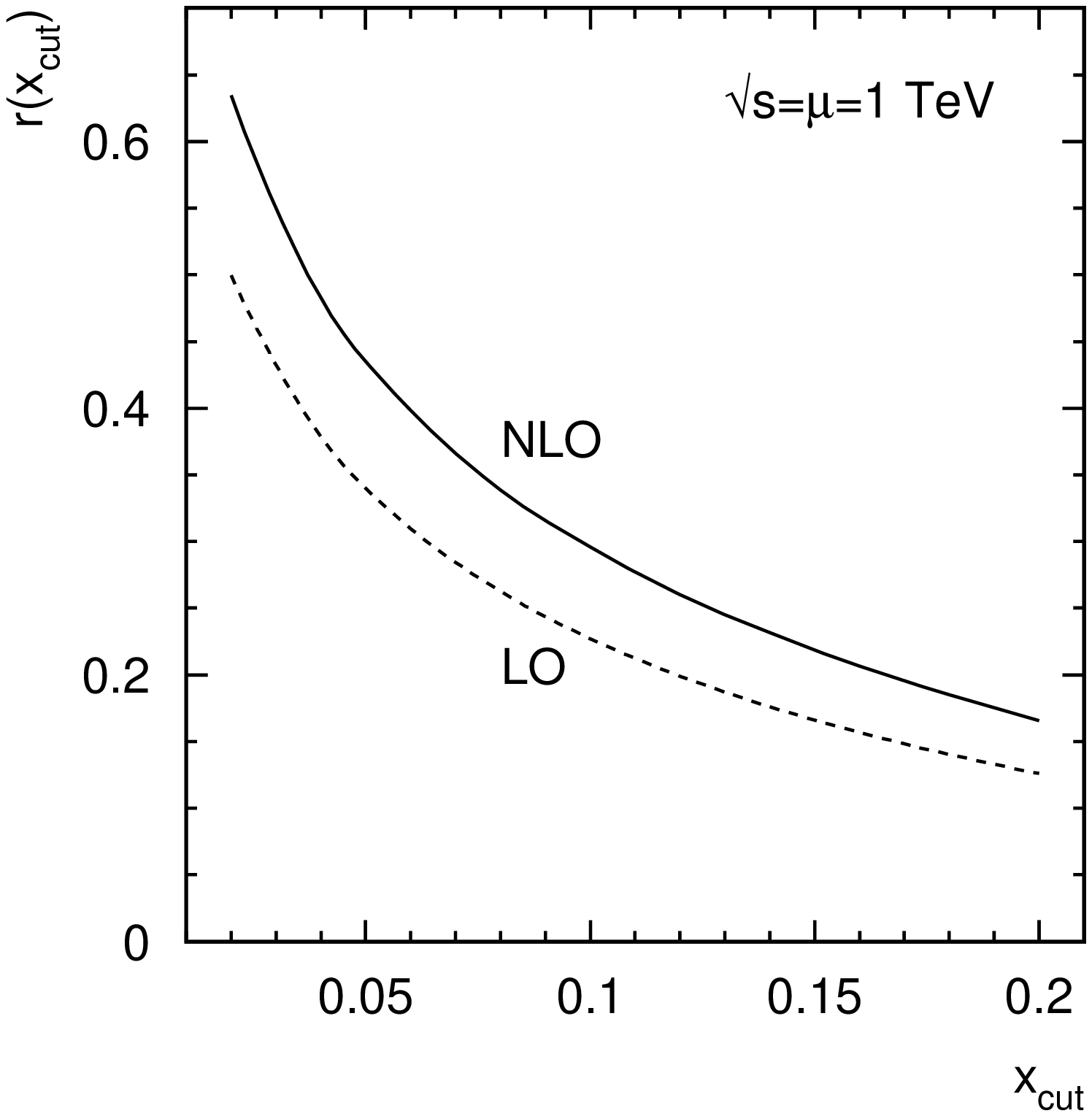,width=9cm,height=9cm}}
\end{picture}
\vskip 0.5cm
\caption{}\label{fig:ycut_1000}
\end{center}
\end{figure}
\begin{figure}[ht]
\unitlength1.0cm
\begin{center}
\begin{picture}(8,8)
\put(0,0){\psfig{figure=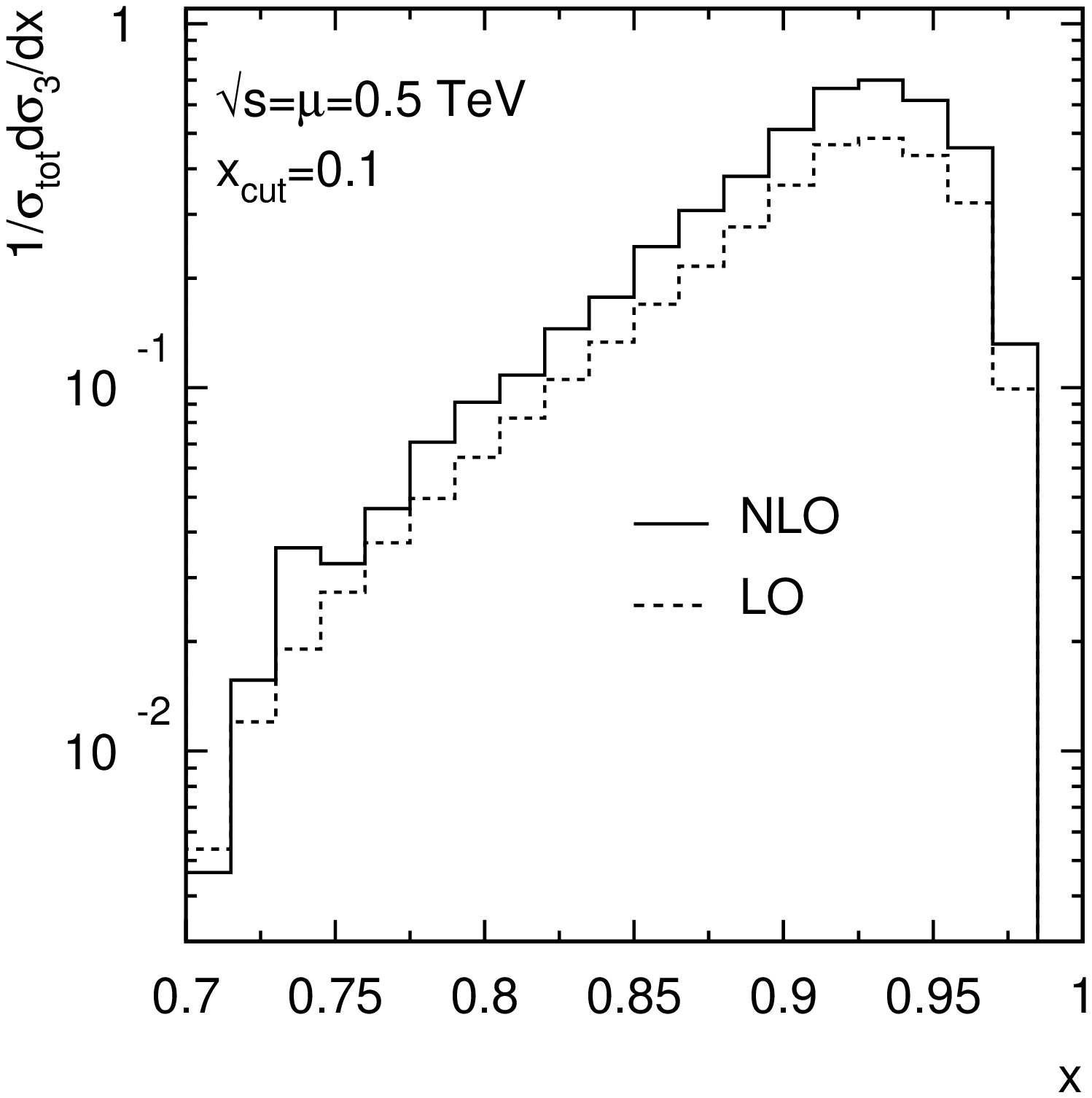,width=9cm,height=9cm}}
\end{picture}
\vskip 0.5cm
\caption{}\label{fig:x}
\end{center}
\end{figure}
\begin{figure}[ht]
\unitlength1.0cm
\begin{center}
\begin{picture}(8,8)
\put(0,0){\psfig{figure=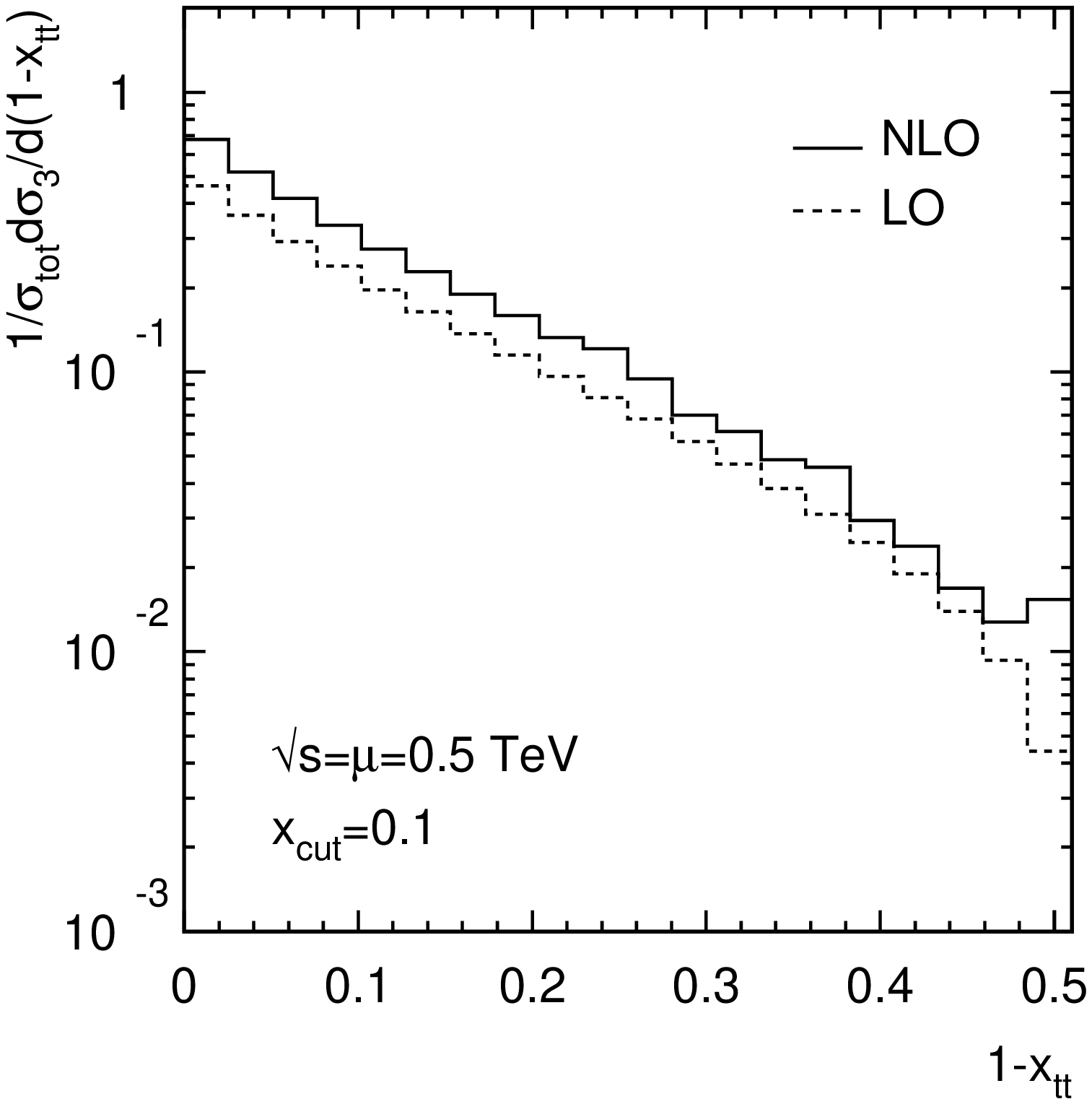,width=9cm,height=9cm}}
\end{picture}
\vskip 0.5cm
\caption{}\label{fig:xg}
\end{center}
\end{figure}
\begin{figure}[ht]
\unitlength1.0cm
\begin{center}
\begin{picture}(8,8)
\put(0,0){\psfig{figure=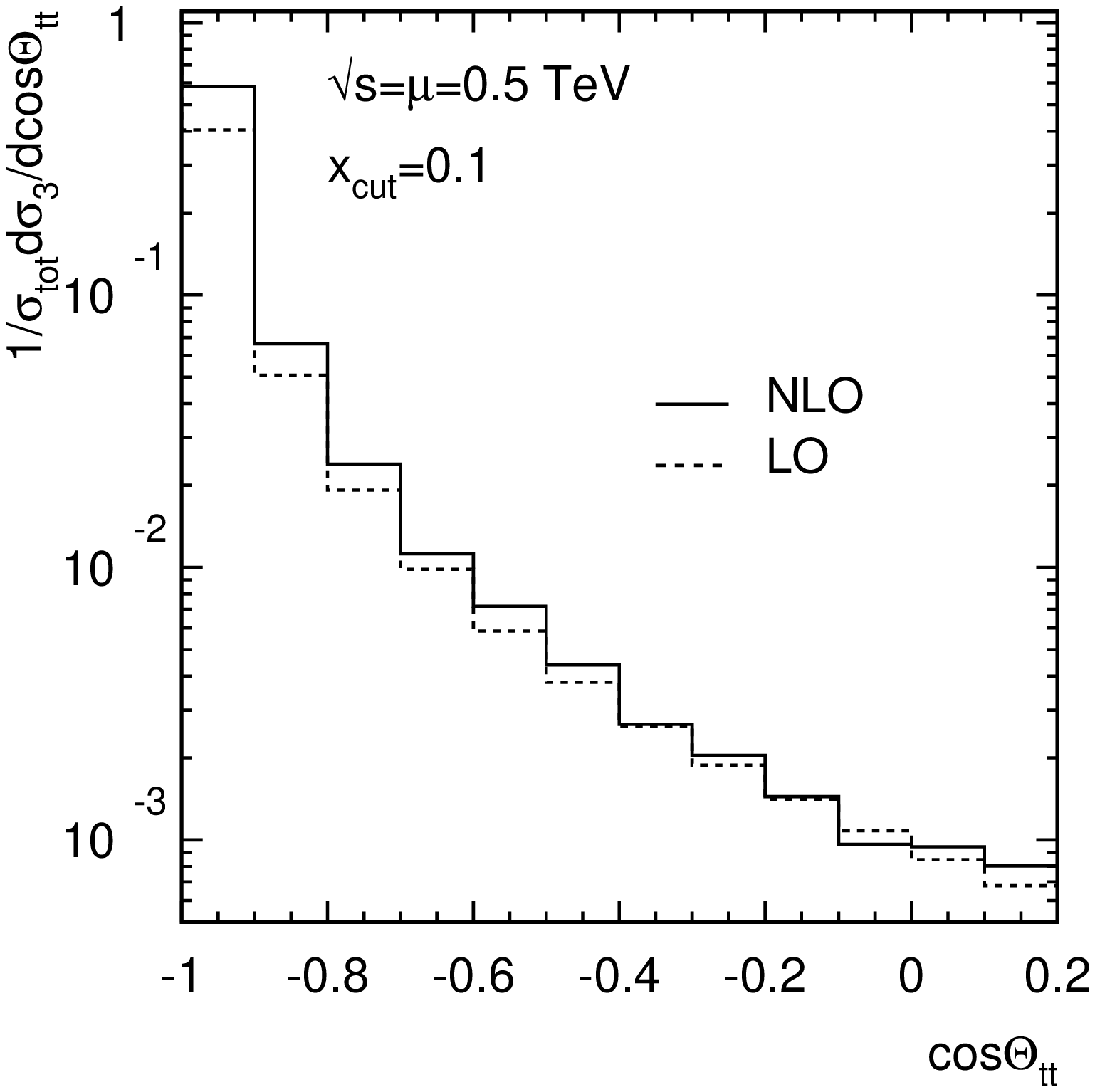,width=9cm,height=9cm}}
\end{picture}
\vskip 0.5cm
\caption{}\label{fig:cos}
\end{center}
\end{figure}
\begin{figure}[ht]
\unitlength1.0cm
\begin{center}
\begin{picture}(8,8)
\put(0,0){\psfig{figure=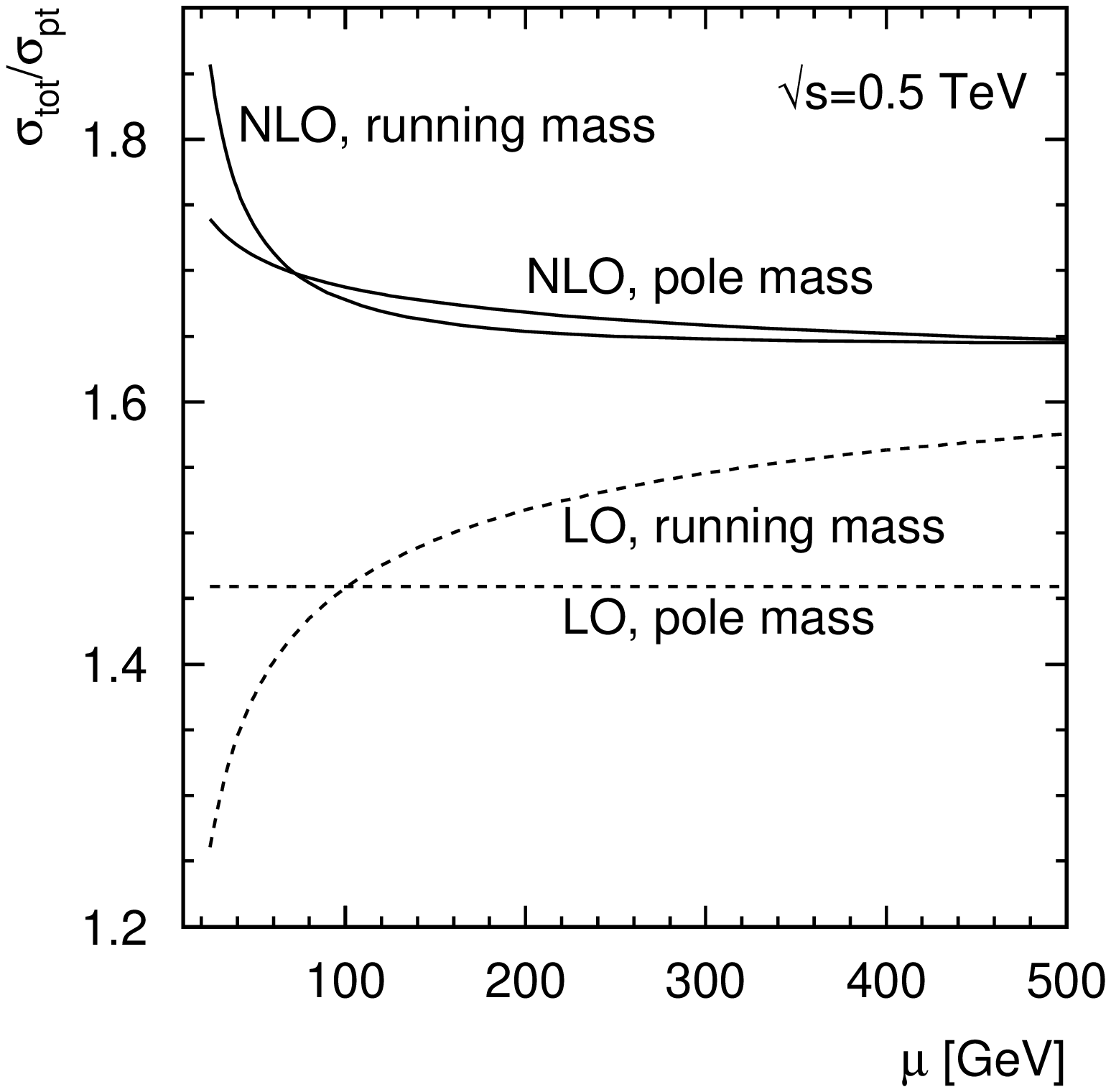,width=9cm,height=9cm}}
\end{picture}
\vskip 0.5cm
\caption{}\label{fig:mudep_tot}
\end{center}
\end{figure}
\begin{figure}[ht]
\unitlength1.0cm
\begin{center}
\begin{picture}(8,8)
\put(0,0){\psfig{figure=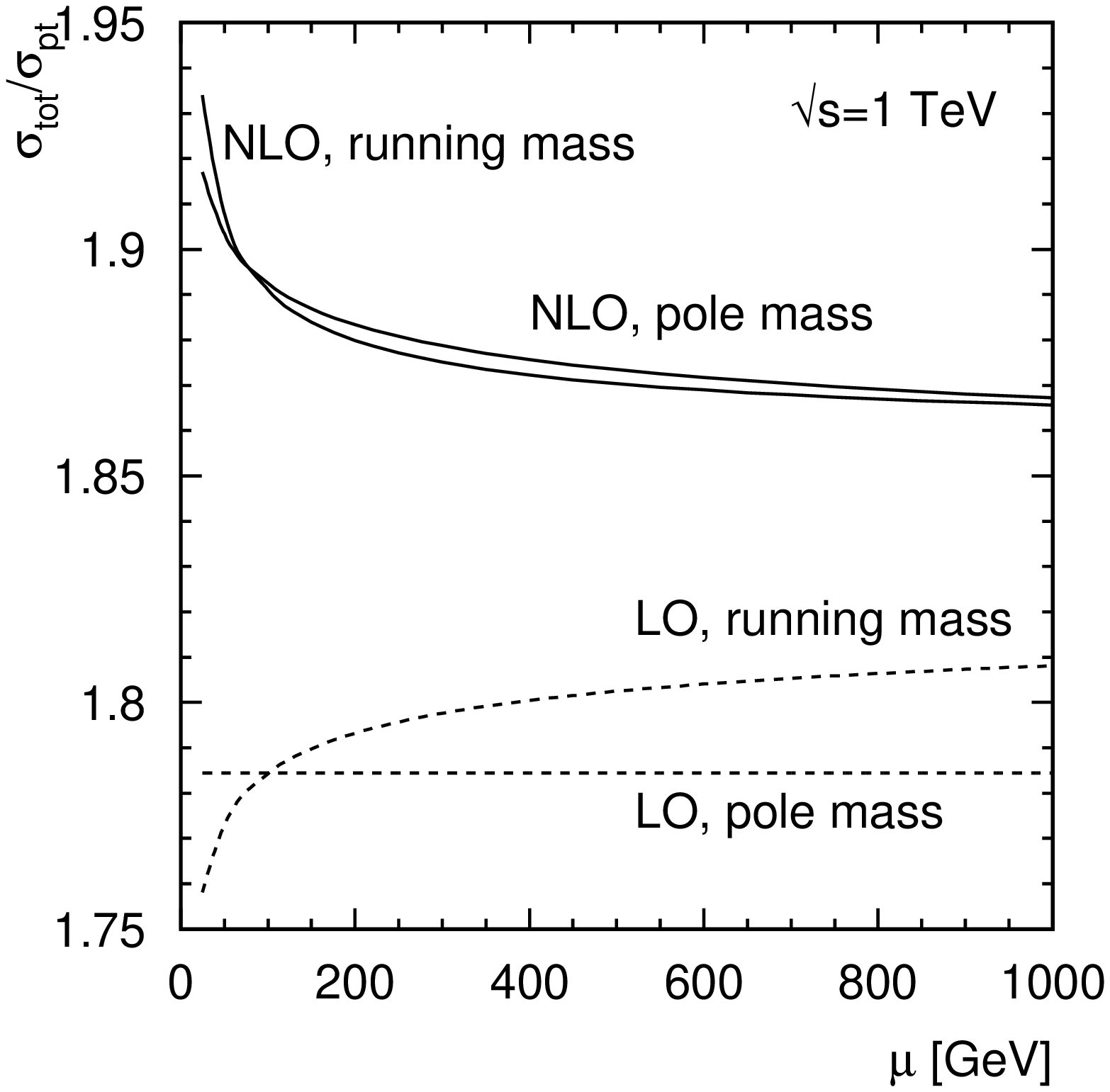,width=9cm,height=9cm}}
\end{picture}
\vskip 0.5cm
\caption{}\label{fig:mudep_1000_tot}
\end{center}
\end{figure}
\begin{figure}[ht]
\unitlength1.0cm
\begin{center}
\begin{picture}(8,8)
\put(0,0){\psfig{figure=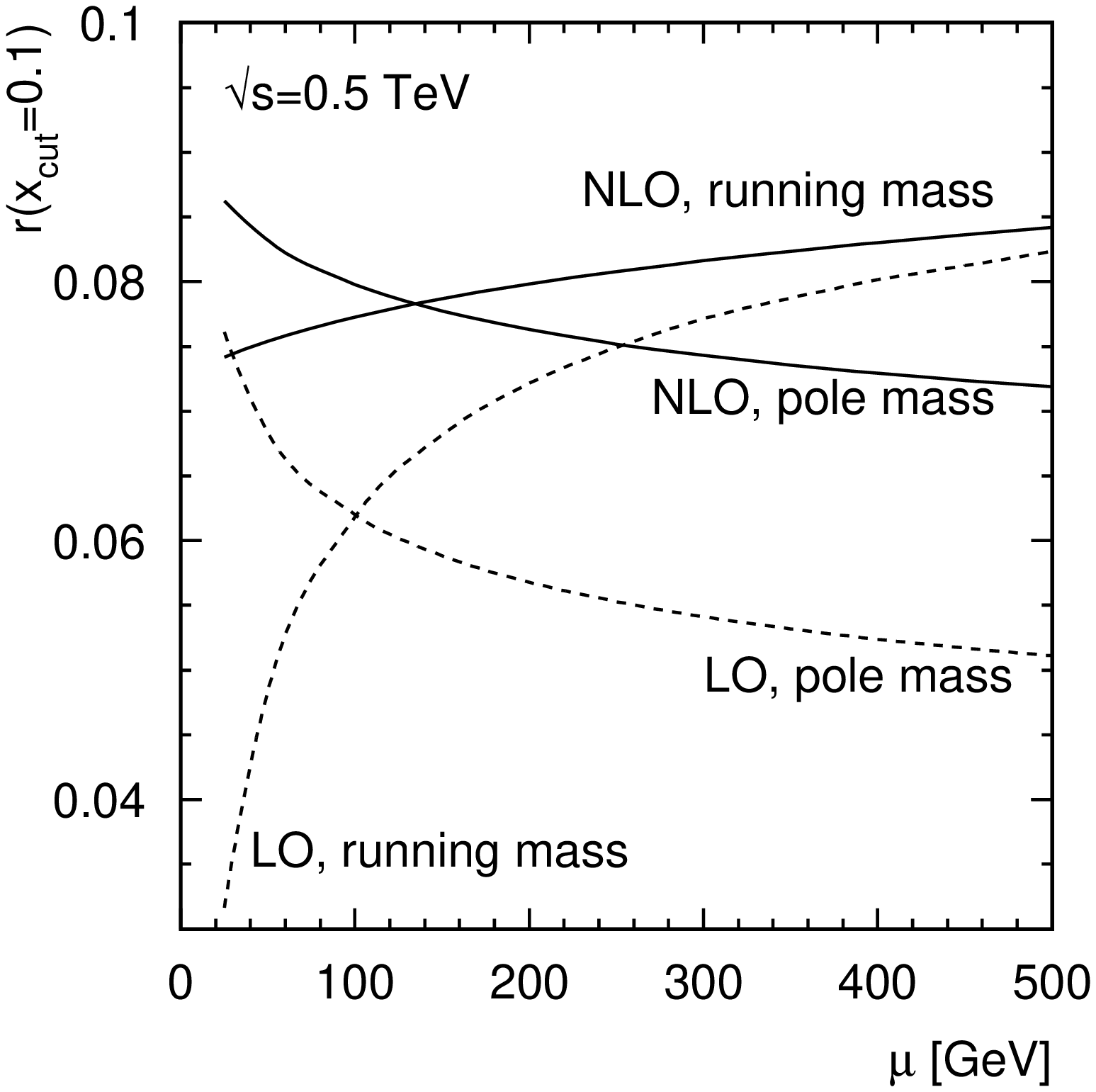,width=9cm,height=9cm}}
\end{picture}
\vskip 0.5cm
\caption{}\label{fig:mudep}
\end{center}
\end{figure}
\begin{figure}[ht]
\unitlength1.0cm
\begin{center}
\begin{picture}(8,8)
\put(0,0){\psfig{figure=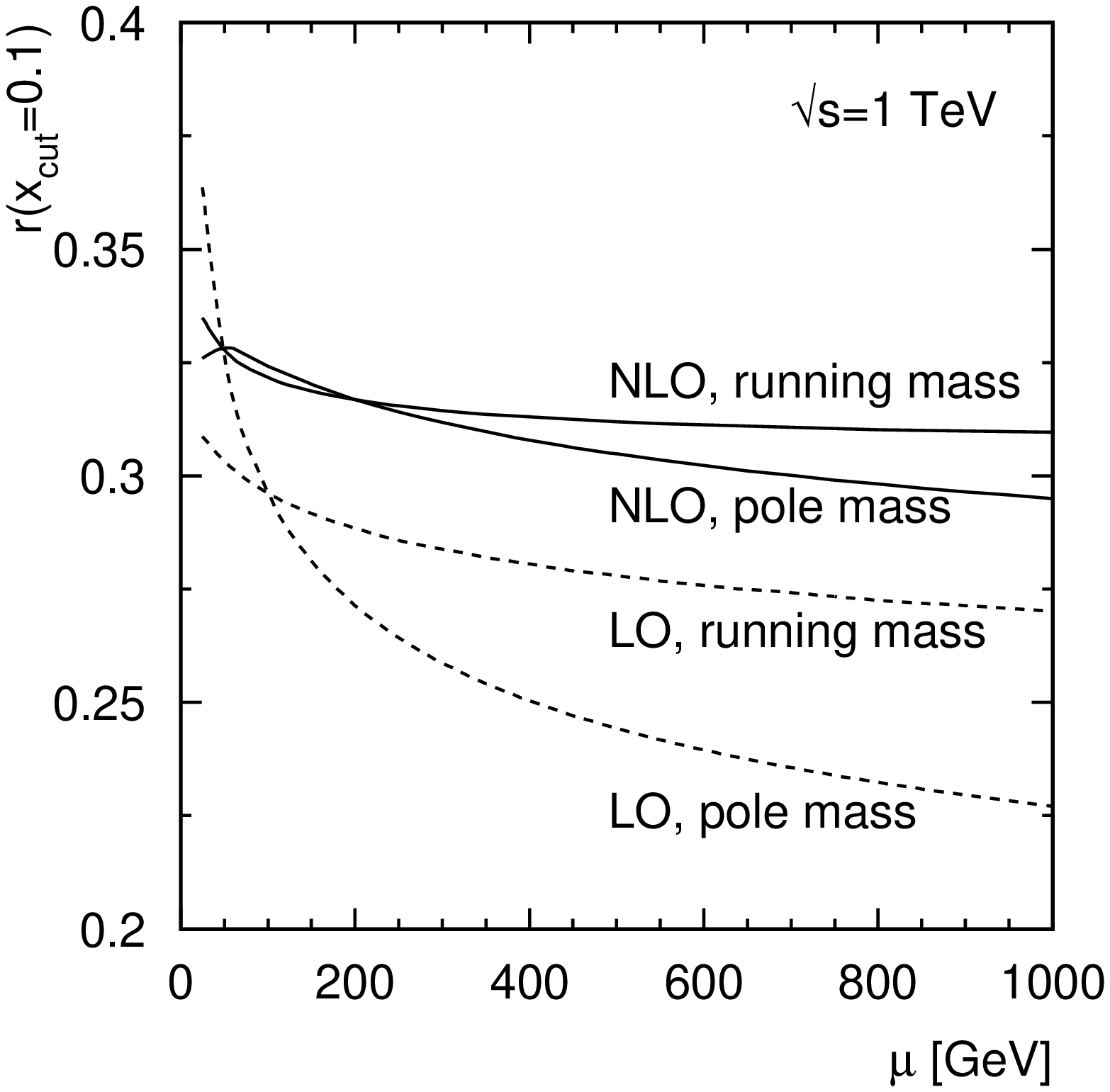,width=9cm,height=9cm}}
\end{picture}
\vskip 0.5cm
\caption{}\label{fig:mudep_1000}
\end{center}
\end{figure}
\begin{figure}[ht]
\unitlength1.0cm
\begin{center}
\begin{picture}(8,8)
\put(0,0){\psfig{figure=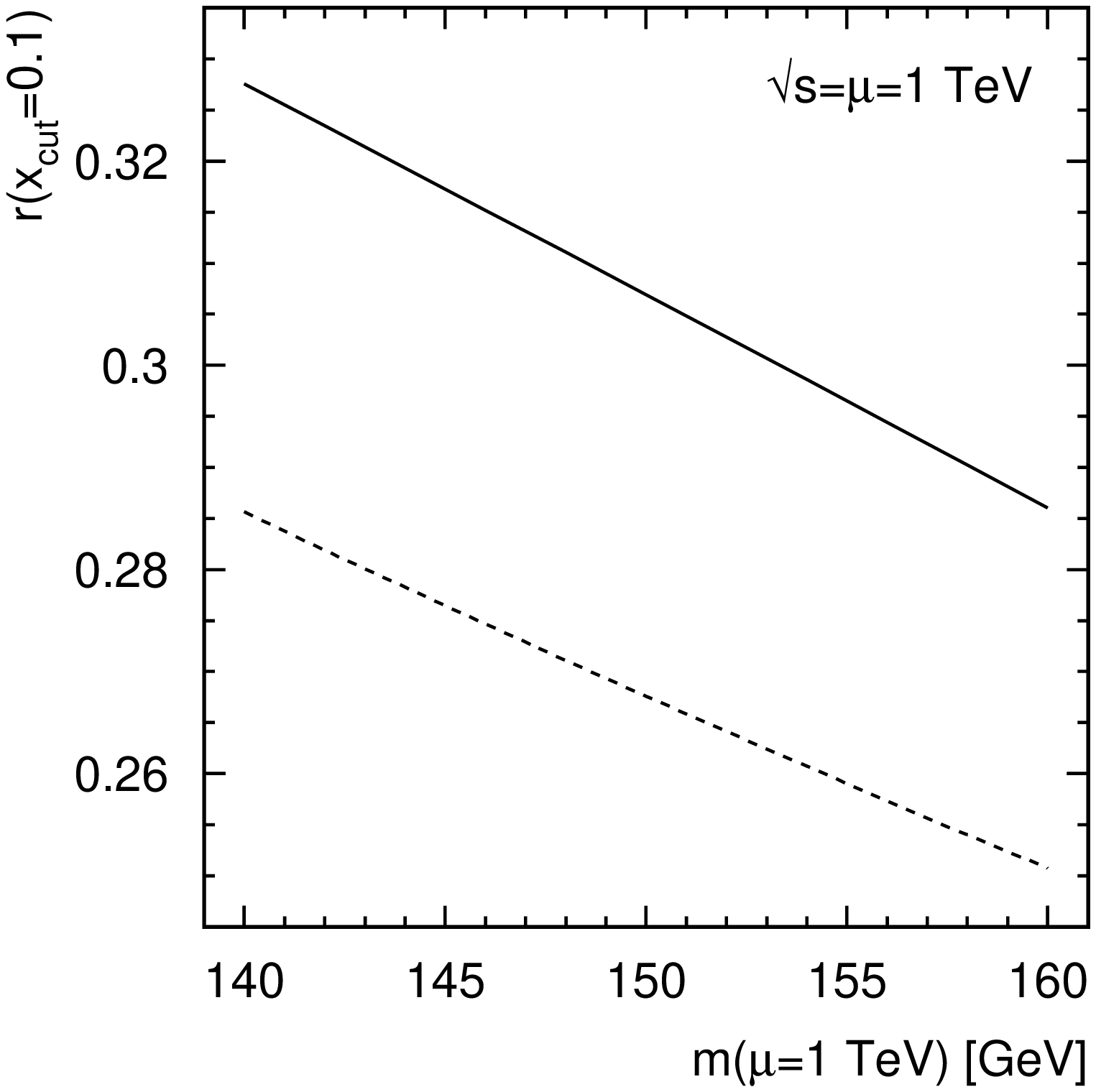,width=9cm,height=9cm}}
\end{picture}
\vskip 0.5cm
\caption{}\label{fig:mass_1000}
\end{center}
\end{figure}
\end{document}